\documentclass{kapproc} 
\usepackage{graphicx}

\normallatexbib

\begin{document}

\articletitle{Point-contact spectroscopy of two-band
superconductor M{\small g}B$_2$ }
\author{I. K. Yanson and Yu. G. Naidyuk}

\affil{B.Verkin Institute for Low Temperature Physics and
Engineering, National Academy  of Sciences of Ukraine, 47 Lenin
Ave., 61103, Kharkiv, Ukraine}

\email{yanson@ilt.kharkov.ua}

\begin{abstract}
The progress in investigation of two-band superconductor MgB$_2$
by the point-contact spectroscopy (PCS) is given. Results of study
of superconducting gap temperature and magnetic field dependence
for two-dimensional $\sigma$ and three-dimensional $\pi$ band and
electron-phonon-interaction spectral function are presented.
Correlation between the gap value and the intensity of the high
T$_c$ driving force -- $E_{2g}$ boron vibration mode, is provided.
PCS data on some nonsuperconducting transition metal diborides are
surveyed for comparison.

\end{abstract}

\begin{keywords}
point-contact spectroscopy, MgB$_2$, two-band/gap
superconductivity, electron-phonon interaction
\end{keywords}

\section{Introduction}

Magnesium diboride, like other diborides MeB$_2$ (Me=Al, Zr, Ta,
Nb, Ti, V etc.), crystalizes in a hexagonal structure, where
honeycomb layers of boron are separated by magnesium ions located
above and below the centers of boron hexagons. The hallmark of
MgB$_2$ is that it becomes superconducting (SC) at high critical
temperature T$_c$ ($\approx $ 40\,K) \cite{Nagamatsu}, which is a
record breaking value among the {\it s-p} metals and alloys. The
outstanding property of MgB$_2$ is that this material represents a
rare example of multi-band (2-D $\sigma$-band and 3-D $\pi$-band)
electronic structure, which are weakly connected with each other.
These bands lead to very uncommon properties. For example, T$_c$
almost does not depend on elastic scattering, unlike for other
two-band superconductors \cite{Mazin}. The maximal upper critical
magnetic field there can achieve a much higher value, than that
for a one-band dirty superconductor \cite{Gurevich}. The
properties of MgB$_2$ have been comprehensively calculated by the
modern theoretical methods, which lead to principal understanding
of their behavior in various experiments.


Electron band structure of MgB$_2$ was calculated in very detail
using different {\it ab initio} methods \cite
{An,Kong,Kortus,Liu,Yildirim}. Topmost is that two filled
incompletely $\sigma$ bands have weak $k_z$ dispersion forming two
nearly cylindrical sheets of the Fermi surface around  $\Gamma $A
$(\Delta )$. Besides they retain their covalent structure
representing unique case of conducting covalent bands which
contribute to the strong electron-phonon coupling. Thus, the hole
branch along $\Gamma $A experiences huge interaction with phonon
$E_{2g}$ mode for carriers moving along $ab$ plane, although its
manifestation is screened effectively by the much faster hole
mobility in $\pi $-bands \cite{Mazin}, which form two 3-D tubular
network. Appropriate electron transport is very anisotropic
($\rho_c/\rho_{ab}\simeq 3.5$ \cite{Eltsev}) with a plasma
frequency (and Fermi velocity) for $\sigma$ band along $c$ axis
being an order of magnitude smaller than that in $ab$ direction
\cite{Brinkman}.

Inelastic $X$-Ray scattering measurements \cite{Shukla}
demonstrated a weakly dispersion branch between 60 and 70 meV in
$\Gamma $A direction with $E_{2g}$ symmetry in $\Gamma $ point.
The linewidth of this mode is about 20$\div $28 meV along $\Gamma
$A direction, while along $\Gamma $M direction it is below the
experimental resolution.  This points to the very strong
electron-phonon interaction (EPI) for this particular lattice
vibration mode.

The SC energy gap distribution on the Fermi surface of MgB$_2$
\cite{Choi} shows maximum gap value along $\Gamma $A direction
which is due to very strong EPI. Just in this direction 2D $\sigma
$ band (cylinders along $\Gamma $A direction) is located. The 3D
$\pi $ band has much smaller EPI, and, correspondingly, the nearly
3 times smaller energy gap. In Ref.\,\cite{Choi} it is shown that
average $\lambda$ value on $\sigma $ band amounts up to $2\div 3$.
Moreover, $\lambda_\sigma $ can be decomposed between different
phonon modes, and it appears that only E$_{2g}$ phonon mode along
$\Gamma $A direction plays a major role with a partial
$\lambda_{\sigma} $ value of about $\simeq 25$ \cite{An1}, though
concentrating in a very restricted phase space.


Driving mechanism for high $T_c$ in MgB$_2$ is connected with the
strong interaction between charge carriers and $E_{2g}$ phonon
modes, corresponding to antiparallel vibration of atoms in the
boron planes.  The electron band structure of MgB$_2$ along
$\Gamma $A direction is such that the Fermi energy of hole
carriers is only 0.5$\div 0.6$\,eV, which shrinks even more while
borons deviate from the equilibrium positions. Together with the
2D structure of the corresponding sheet of the Fermi surface, this
leads to constant density of states at the Fermi energy and,
correspondingly, to very strong EPI. Cappelluti {\it et al.}
\cite{Cappelluti} point out that the small Fermi velocity for
charge carriers along $\Gamma $A direction leads to large
nonadiabatic correction to $T_c$ (about twice as much compared
with adiabatic Migdal-Eliashberg treatment). Although this
interaction is a driving force to high $T_c$ in this compound, it
does not lead to crystal structure instability, since it occupies
only a small volume in the phase space.

According to theoretical models $\pi $ and $ \sigma $ bands  in
MgB$_2$ are weakly connected. However, the energy gap of $\pi $
band goes to zero at the same $T_c$ as in the bulk, and
correspondingly the $2\Delta _\pi (0)/kT_c=1.4$, which is much
less than the weak coupling BCS theory predicts. One can think of
$\pi $ band as having intrinsically much lower $T_c\approx 10$\,K
than the bulk \cite{Bouquet} and at higher temperatures its
superconductivity is induced by proximity effect in the {\bf
k}-space from $\sigma $ band \cite{YansonPRB}. This proximity
effect is very peculiar. From one side, this proximity is induced
by the interband scattering between $\pi $ and $\sigma $ sheets of
the Fermi surface. On the other, the charge carriers connected
with $\pi $ band are mainly located along the magnesium planes,
which can be considered as a proximity effect in the coordinate
space for alternating layers of $S-N-S$ structure, although at
microscopically scale. Thus, MgB$_2$ is a good example to study
crossover between two-band superconductivity and simple proximity
effect structure.

\section{Samples and measurements}

In this work the results for two kind of samples are surveyed. The
first is  thin c-axis oriented films with the thickness of about
several hundreds of nanometer \cite{Sung-Ik}. The residual
resistance is about several tens of $\mu \Omega \,$cm with
residual resistance ratio (RRR) $ \simeq 2.2$ pointing out that
films have a disorder between crystallites. It does not exclude
that on some spots the films contain clean enough small single
crystals on which we occasionally may fabricate a point contact
(PC). Normally, the contacts were prepared by touching the film
surface by noble metal counter electrode (Cu, Au, Ag) in the
direction perpendicular to the substrate. Thus, nominally the
preferential current direction in PC is along {\it c} axis.
Nevertheless, since the surface of the films contains terraces
with small crystallites, PC to {\it ab} plane of these
crystallites is also possible.

The second type of samples are single crystals \cite{Lee}.
Crystals are plate-like (flakes) and have sub-millimeter size.
They were glued by silver epoxy to the sample holder by one of
their side faces. The opposite face of flakes was used as a
"needle" to gently touch the noble metal counter electrode in
liquid helium. In this way we tried to make preferentially a
contact along {\it ab} plane. In average, in the bulk, the single
crystals are cleaner than the films, but one should be cautious,
since the properties of the crystal surface differ from the
properties of the bulk, and fabrication of PC may introduce
uncontrolled further defects in the contact area.

Thus, {\it a priori} one cannot define the structure and
composition of the obtained contacts. Nevertheless, much of that
issue can be said by measuring various characteristics of a
contact. Among those the most important is the
Andreev-reflection-non-linearities of the $I-V$ curves in the SC
energy-gap range. The magnetic field and temperature dependencies
of the SC non-linearities supply us with additional information.
And finally, much can be extracted from the $I-V$ nonlinearities
in the normal state (so called, PC spectra). The more information
about the electrical conductivity at different conditions of the
particular contact we can collect, the more detailed and defined
picture of it emerges. It is not an easy task, since a contact has
a limited life time, due to electrical and mechanical shocks.

Let us give a rough estimation of the distance scales involved in
the problem. The crystallite size of films is of the order of
100\,nm (see \cite{Sung-Ik}). The contact size $d$ in ballistic
regime equals $d\simeq\sqrt{\rho l/R}$ (the Sharvin formula).
Taking $\rho l\cong 0.7\times 10^{-6}\Omega \,$cm$\times
7\,10^{-6}$cm$\cong 0.5\times 10^{-11} \Omega \,$cm$^2$
\cite{Eltsev}, we obtain $d\simeq 7$ nm both along $ab$ and $c$
directions for typical resistance of 10 $\Omega$. If we suppose
that a grain is dirty (with very short mean free path), then we
apply the Maxwell formula $d\sim\rho/R$ with the results for $d$
about 0.7\,nm and 2.6\,nm for $ab$ and $c$ directions,
respectively, taking $\rho$ for corresponding directions from the
same reference \cite{Eltsev}. Thus, the contact size can be of the
order or smaller than the electronic mean free path ($l_{ab}=70$
nm and $l_c=18$\,nm, according to \cite{Eltsev}), which means that
we are working in the spectroscopic regime, probing only a single
grain.

Rowell \cite{Rowell}, analyzing a big amount of experimental data
for resistivity and its temperature dependence, came to the
conclusion that for highly resistive samples only a small part of
the effective cross section should be taken into account. The
reason is that the grains in MgB$_2$ are disconnected by oxide of
magnesium and boron to great extent. For PCS previous analysis
leads us to the conclusion that the contact resistance is
frequently measured only for a single grain, either for several
grains, with their intergrain boundaries facing the contact
interface. This is due to the current spreading with the scale of
the order of the contact size $d$ near the constriction.

\section{Theoretical background of PCS}


The non-linearities of the $I-V$ characteristic of a metallic
contact, when one of the electrodes is in the SC state, can be
written as \cite {Khlus}

\begin{equation}
I\left( V\right) \simeq  V/R_0-\delta I_{ph}^N(V)+I_{exc}(V)
\label{I-V}
\end{equation}
Here $R_0$ is the contact resistance at zero bias in the normal
state. $\delta I_{ph}^N(V)$ is the backscattering inelastic
current which depends on the electron mean free path (mfp) $l$.
For the ballistic contact this term amounts to
\begin{equation}
\delta I_{ph}^N(V)\sim  (d/l_{in})I(V)  \label{inelastic}
\end{equation}
where $l_{in}$ is the inelastic electron mfp, and $d$ is the
characteristic contact diameter. If the electron flow through the
contact is diffusive ($ l_{el}\ll d$, $l_{el}$ being an elastic
mfp) but still spectroscopic, since $ \sqrt{l_{in}l_{el}}\gg d$,
then the expression (\ref{inelastic}) should be multiplied by
$l_{el}/d$. This decreases the characteristic size, where the
inelastic scattering being essential, from $d$ to $l_{el}$
($d\rightarrow l_{el}$), and for short $l_{el}$ makes the
inelastic current very small. We notice that the inelastic
backscattering current $\delta I_{ph}^N(V)$ in the SC state is
approximately equal to the same term in the normal state. Its
second derivative turns out to be directly proportional to the EPI
function $\alpha^2(\omega)\,F(\omega)$ \cite{KOS,YansonSC}
\begin{equation}
\label{pcs} -d^2I/dV^2\simeq (8\,ed/3\,\hbar v_{\rm
F})\alpha^2(\omega)\,F(\omega)
\end{equation}
where $\alpha$ describes the strength of the electron interaction
with one or another phonon branch, $F(\omega)$ stands for the
phonon density of states. In PC spectra the EPI spectral function
$\alpha^2(\omega)\,F(\omega)$ is modified by the transport factor,
which increases strongly the backscattering processes
contribution.

In the SC state the excess current $I_{exc}$\,(\ref{I-V}), which
is due to the Andreev reflection of electron quasiparticles from
the $N-S$ boundary in a $N-c-S$ contact ($c$ stands for
''constriction''), can be written as

\begin{equation}
I_{exc}\left( V\right) =I_{exc}^0+\delta I_{exc}\left( V\right)
\label{excess}
\end{equation}
where $I_{exc}^0\approx \Delta /R_0\approx const$ for $eV>\Delta $
($\Delta $ being the SC energy gap).

Nonlinear term in the excess current (\ref{excess}) in its turn
can be decomposed in two parts, which depend in a different way on
the elastic scattering of electron quasiparticles:

\begin{equation}
\delta I_{exc}\left( V\right) =\delta I_{exc}^{el}\left( V\right)
+\delta I_{exc}^{in}\left( V\right)  \label{excess1}
\end{equation}
where $\delta I_{exc}^{el}\left( V\right) $ is of the order of
$\left( \Delta /eV\right) I_{exc}^0$, and $\delta
I_{exc}^{in}\left( V\right) \sim \left( d/l_{in}\right)
I_{exc}^0$. Notice, that the latter behaves very similar to the
inelastic backscattering current $\delta I_{ph}^N(V)$, namely, it
disappears if $l_{el}\rightarrow 0$, while the first term in the
right hand side of expression (\ref{excess1}) does not depend in
the first approximation on $l_{el}$. This enables to distinguish
elastic term from inelastic. Finally, all excess current terms
disappear when destroying the superconductivity, while $\delta
I_{ph}^N(V)$ remains very similar in both SC and normal states.

From the expressions (\ref{I-V}), (\ref{inelastic}),
(\ref{excess}) and (\ref {excess1}), it becomes clear that only on
the relatively {\it clean} spots, one can observe the inelastic
backscattering current $\delta I_{ph}^N(V)$ provided the excess
current term $\delta I_{exc}^{in}\left( V\right) $ is negligible.
The latter can be cancelled by suppression of superconductivity
either with magnetic field or temperature. On the contrary, in the
SC state, for {\it dirty} contacts, all the inelastic terms are
very small, and the main non-linearity is provided by the $\Delta
(eV)$-dependence of the excess current.


Brinkman {\it et al.} \cite{Brinkman} have shown that even along
{\it ab}-plane the contribution of $\sigma $ band for MgB$_2$ is
less than that of $\pi $ band, to say nothing of the direction
along {\it c} axis, where it is negligible small. The calculation
predicts that if the ''{\it tunneling cone}'' is about several
degrees from precise {\it ab} plane, then two SC gaps should be
visible in tunneling characteristics. In other directions only a
single gap, corresponding to $\pi $ band, is visible. We will see
below that this prediction is fulfilled in PC experiment, as well.

Even worse the things are when one tries to measure the
anisotropic Eliashberg function by means of the SC tunneling. The
single-band numerical inversion program gives  non-certain
results, as was shown in Ref.\,\cite{Dolgov}.

Point-contact spectroscopy in the normal state can help in this
deadlock situation. It is known that the inelastic backscattering
current is based on the same mechanism as an ordinary homogeneous
resistance, provided the maximum  energy of charge carriers  is
controlled by applied voltage. The electrical conductivity of
MgB$_2$ can be considered as parallel connections of two channels,
corresponding to $\pi $ and $ \sigma $ bands \cite{Mazin}. The
conductivity of $\pi $ band can be blocked by Mg-atoms disorder.
This situation is already obtained in experiment, when the
temperature coefficient of resistivity increases simultaneously
with increase of residual resistivity, which leads to the
violation of the Matthiessen's rule (see  Fig.\,3 in
\cite{Mazin}). In this case we obtain the direct access to $\sigma
$-band conductivity, and the measurements of PC spectra of EPI for
$\sigma$ band are explicitly possible in the normal state. Below
we will see that this unique situation happens in single crystals
along {\it ab} plane.

\section{Experimental results}

\subsection{Superconducting energy gaps}

\begin{figure}
\includegraphics[width=8cm,angle=0]{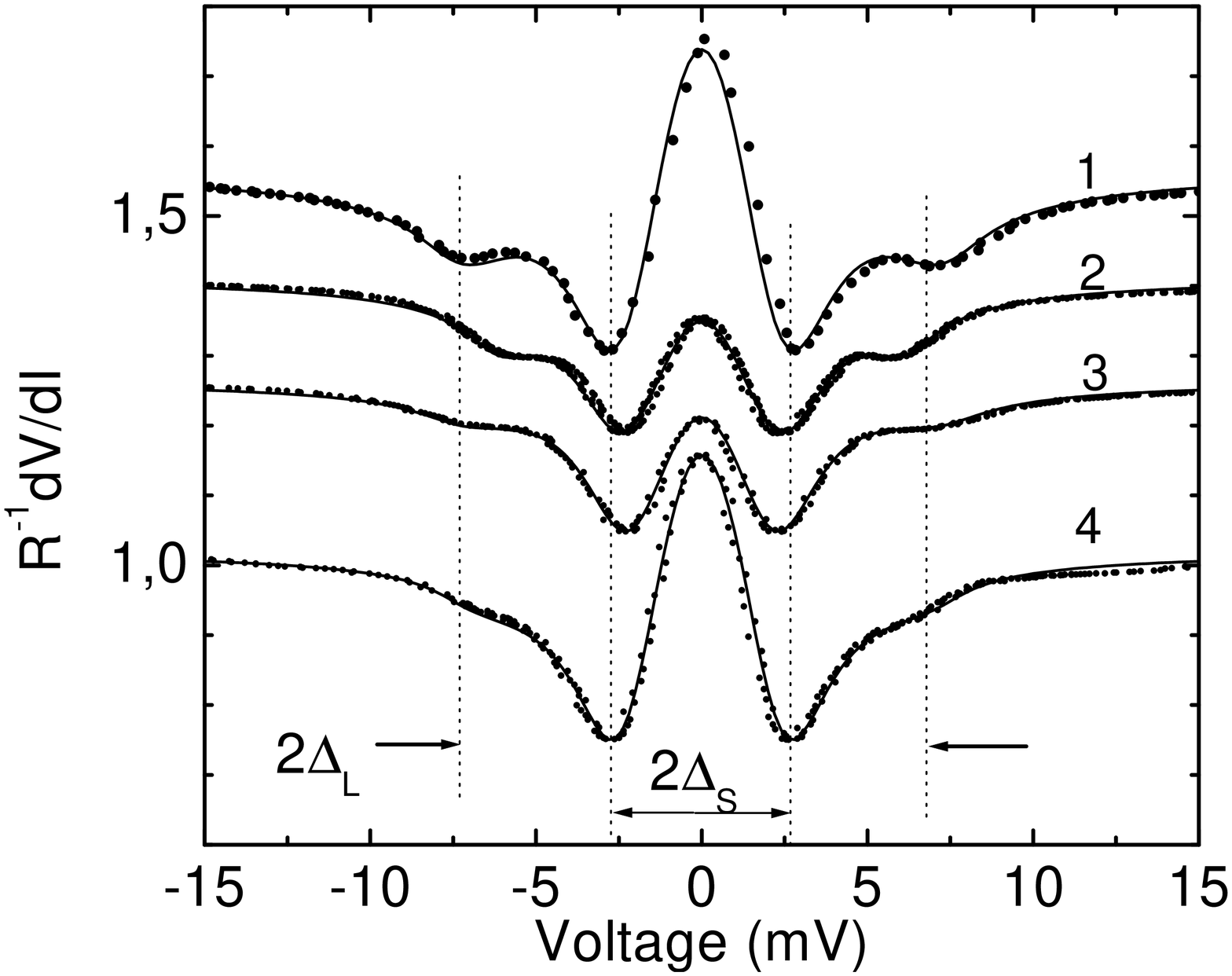}
 \vspace{-4cm}
\narrowcaption{$dV/dI$ (experimental dots) for 4 contacts between
MgB$_2$ thin film and Ag with corresponding BTK fitting (lines)
\cite{Naidyuk1}. $\Delta_{L(S)}$ stands for large (small) SC
energy gap. After Naidyuk  et~al.~ \cite{Naidyuk1}.}
\label{MgB2del}
\end{figure}
\begin{figure}
\includegraphics[width=8cm,angle=0]{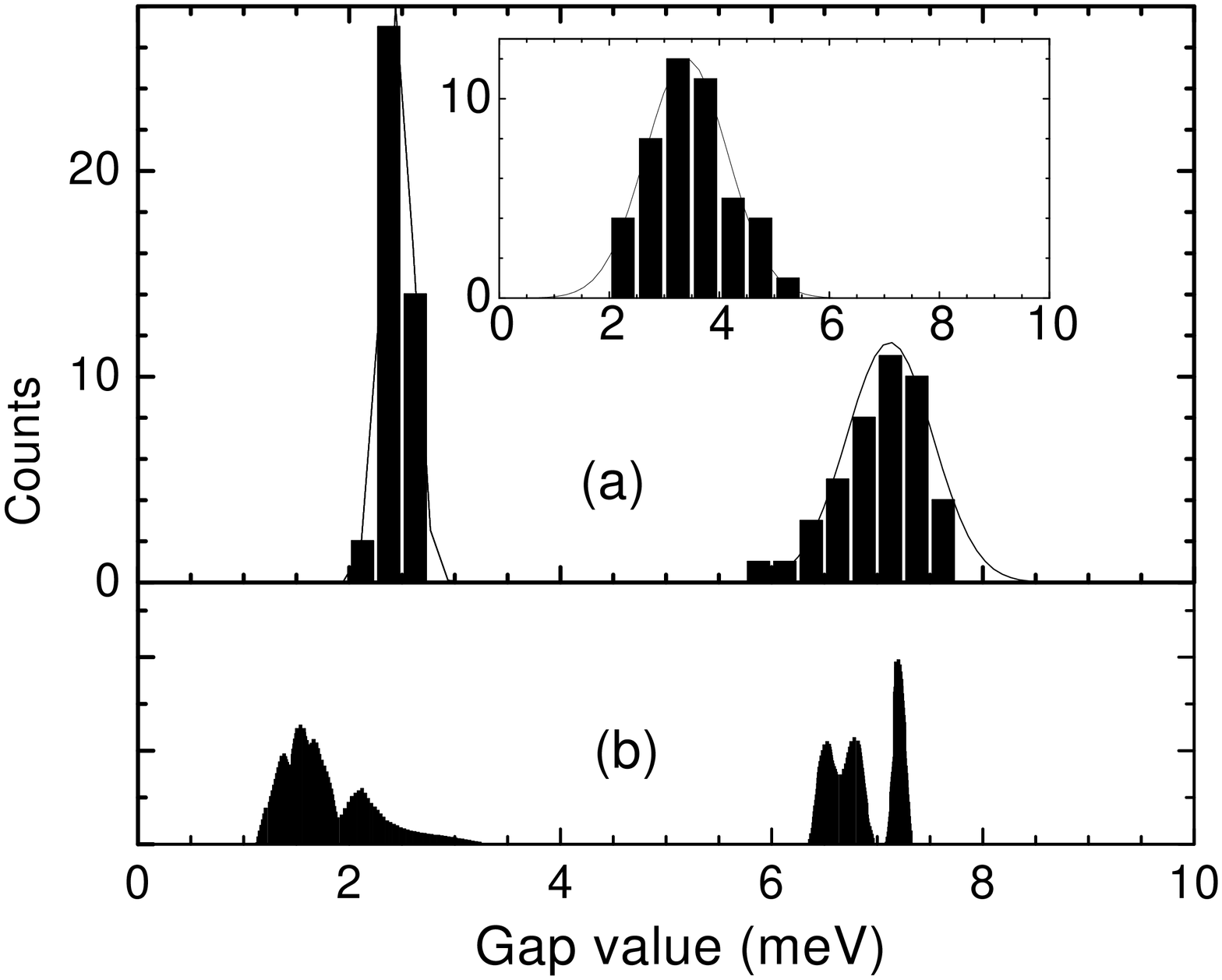}
 \vspace{-4.5cm}
 \narrowcaption{(a) Superconducting energy gap distribution
for about hundred different junctions prepared on MgB$_2$ film
\cite{Naidyuk1}. Inset shows single gap distribution. (b)
Theoretical gap distribution after \cite{Choi}.} \label{MgB2hist}
\end{figure}

Typical shapes of $dV/dI$ with  Andreev reflection features are
shown in Fig.\,\ref{MgB2del}.
$dV/dI$ curves exhibit two sets of energy gap minima distributed
as shown in  Fig. \ref{MgB2hist} (upper panel), at 2.4$\pm $ 0.1
and 7.1$\pm $0.4 meV. These curves are nicely fitted by BTK
\cite{BTK} theory (with a small $\Gamma$ parameter) for two
conducting channels with an adjusted gap weighting factor
\cite{Naidyuk1}. The second kind of $dV/dI$ represents only one
gap structure and is better fitted with a single gap provided an
increased depairing parameter $\Gamma $ (Fig.\,\ref{MgB2hist},
inset). According to the calculation in \cite{Brinkman} strong
impurity scattering will cause the gaps to converge to $\Delta
\simeq$4.1\,meV and $T_c$ to 25.4\,K. Therefore the single gap
spectra reflect a strong interband scattering due to impurities,
which likely causes a "semiconducting-like" behavior of $dV/dI$
above T$_c$. These two kinds of gap structure constitute about
equal parts of a total number of about hundred junctions. Usually
the contribution of the large gap in the double-gap spectra is an
order of magnitude lower than that of the small one, which is in
line with small contribution of $\sigma$ band in conductivity
along c axis \cite{Brinkman}.

In the lower panel of Fig.\,\ref{MgB2hist} the theoretical
prediction of energy gap distribution \cite{Choi} is shown. One
can see that the theoretical positions of distribution maxima
approximately coincide with the experimental values. Only the
low-lying maximum is not seen in the experiment.


The same variety of energy gap structure is observed for single
crystals as well, but with some peculiarity due to preferential
orientation along {\it ab} plane. The most amazing of them is the
observation of a $dV/dI$ gap structure in \cite{Naidyuk3} with
visually only the larger gap present. Such kind of spectra were
not observed in thin films. It means that the conductivity is
governed only by $\sigma $ band. This may be caused that $\pi $
band is blocked completely by Mg disorder or by oxidation of Mg
atoms on {\it ab}-side surface of the crystal. At the same time,
in single crystal there is much less scattering in the boron
planes, due to the robustness of B-B bonds. We will see below that
just this case enables us to observe directly the most important
$E_{2g}$ phonon mode in the electron-phonon interaction within
$\sigma$ band.
\begin{figure}
\includegraphics[width=9cm,angle=0]{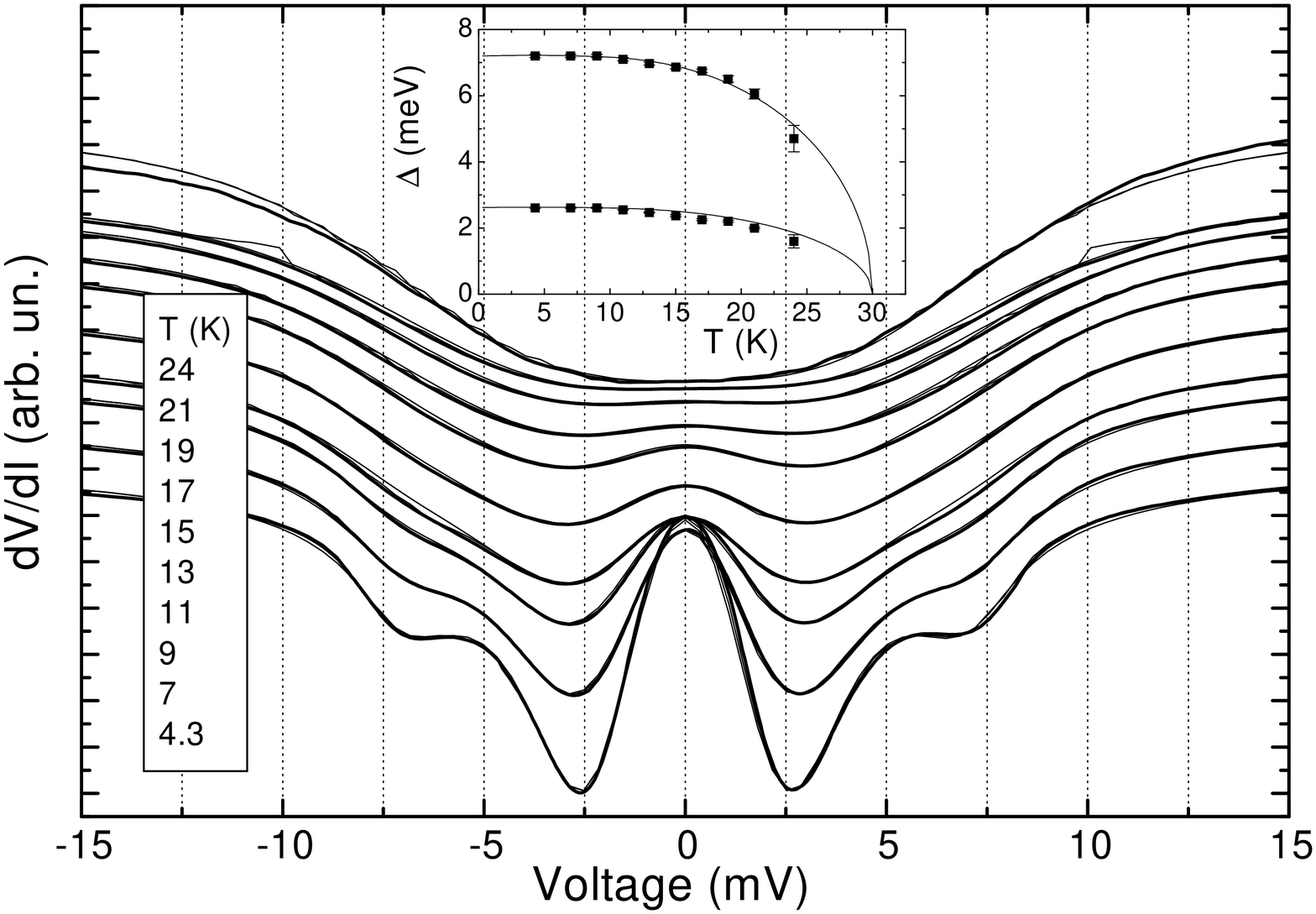}
\caption[]{$dV/dI$ curves (solid lines) at different temperature
for the same as in Fig.\,\ref{M163m} junction with their BTK
fittings (thin lines). Inset: Temperature dependencies of large
and small SC energy gaps obtained by BTK fitting. Solid lines
represent BCS-like behavior.} \label{M163t}
\end{figure}

\begin{figure}
\includegraphics[width=9cm,angle=0]{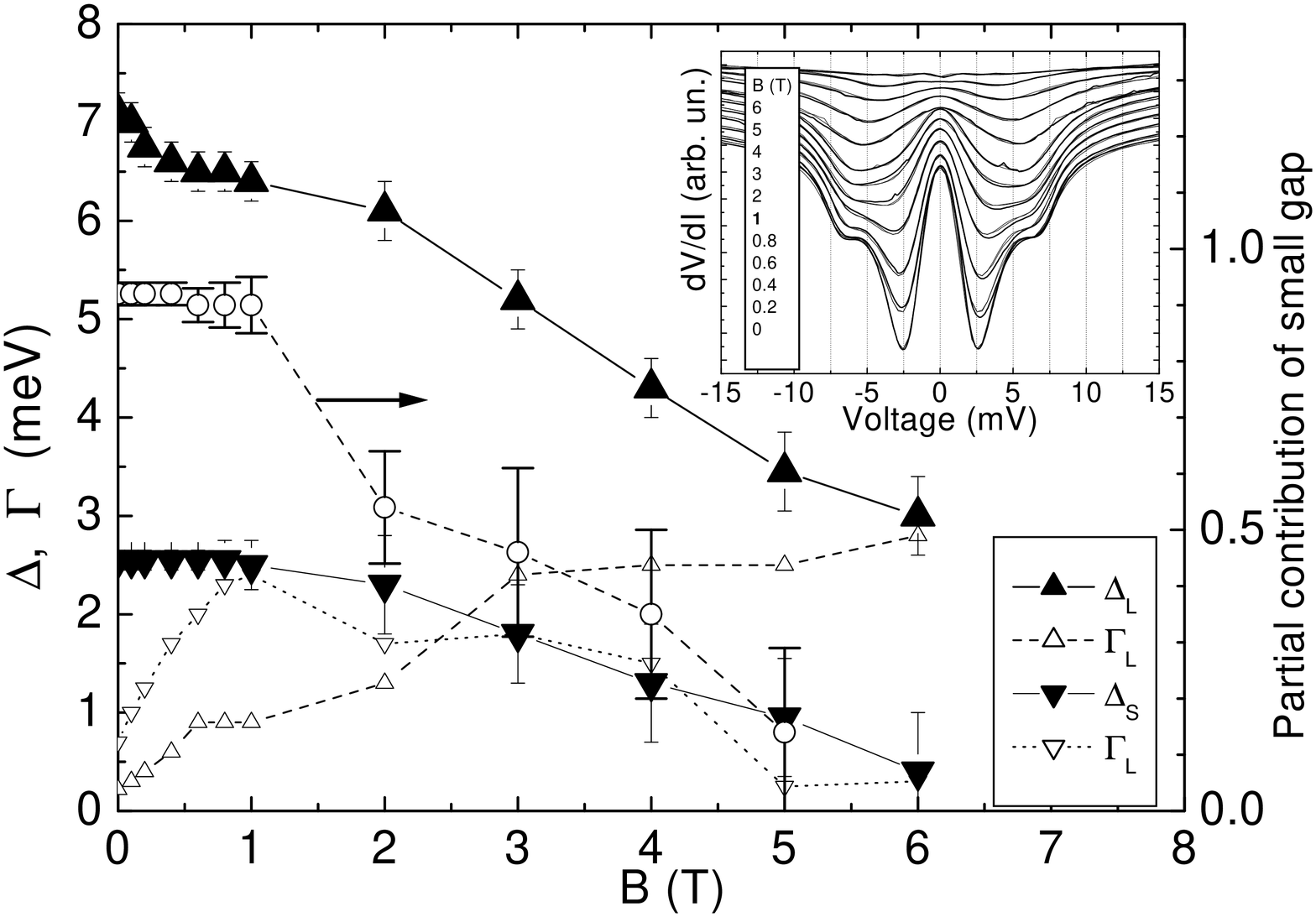}
\caption[]{Magnetic field dependencies of large and small SC
energy gaps (solid triangles) obtained by BTK fitting of the
$dV/dI$ curves from inset. Open triangles show $\Gamma$ value for
large and small gap, correspondingly. Circles demonstrate
depressing of small gap contribution to the $dV/dI$ spectra by
magnetic field. The lines connect the symbols for clarity. Inset:
$dV/dI$ curves (solid lines) at different magnetic field for the
single crystal MgB$_2$-Cu 2.2\,$\Omega$ junction along $ab$ plane
with their BTK fittings (thin lines). } \label{M163m}
\end{figure}

Figs.\,\ref{M163t} and  \ref{M163m} display the series of
temperature and magnetic field dependencies of the $dV/dI$ curves
with their BTK fittings, respectively. Here, at low field
(temperature) the two separate sets of the gap minima are clearly
seen. The temperature dependence of both gaps follows the BCS
prediction (see inset in Fig.\,\ref{M163t}). For temperatures
above 25\,K their behavior is unknown because this particular
contact did not survive during the measurements likely due to
thermal expansion of sample holder.

Fig.\,\ref{M163m} shows magnetic field dependencies of large and
small gaps. Surprisingly, the small gap value is not depressed by
field about 1\,T, and the estimated critical field about 6\,T is
much higher as stated in \cite{Gonnelli,Samuely}, although the
intensity of small gap minima is suppressed rapidly by a field
about 1\,T. Correspondingly, small gap contribution $w$ ($w$
inversely depends on $\Gamma$ value, therefore nearly constant $w$
value between 0 and 1\,T is due to the fact that $\Gamma$ rises by
factor 4 at 1\,T) to the $dV/dI$ spectra decreases by magnetic
field significantly from 0.92 to 0.16 (see Fig.\,\ref{M163m}),
while $w$ versus temperature even slightly increases from 0.92 at
4.3\,K to 0.96 at 24\,K (not shown). Theoretical investigation of
the field dependence of maximum pair potential in two band
superconductor MgB$_2$ by Koshelev and Golubov \cite{Koshelev}
shows that for both gaps critical field is the same. Additionally,
in recent experimental publication Bugoslavsky {\it et al.}
\cite{Bugoslavsky} reported that both order parameters survive to
a common magnetic field, while Gonnelli {\it et al.}
\cite{Gonnelli1} corrected their previous claims and mentioned
that identification of magnetic field at which the $\pi$ band
features in $dV/dI$ visually disappears with critical field for
the $\pi$ bang might not be correct.

\subsection{Electron-Phonon interaction}

\subsubsection{PC EPI spectra of non-SC diborides}
\begin{figure}
\includegraphics[width=12cm,angle=0]{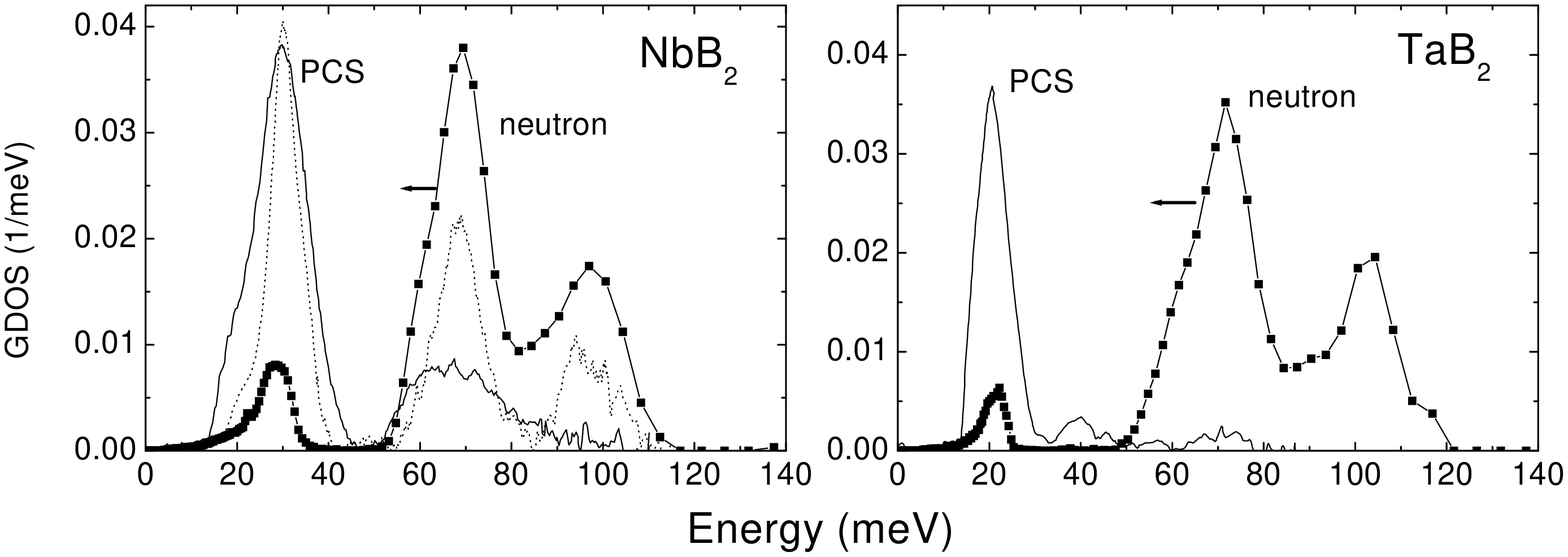}
 \vspace{-4cm}
\caption[]{Comparison of phonon DoS neutron measurements
\cite{Heid} (symbols) with PC spectra for NbB$_2$ and TaB$_2$
\cite{Naidyuk2} after subtracting of rising background (solid
curves). Dotted curve in left panel shows PC spectrum for ZrB$_2$
\cite{Naidyuk2} for confrontation.} \label{Heid}
\end{figure}

In Fig.\,\ref{Heid} the PC EPI spectra  $ d^2V/dI^2 \propto
-d^2I/dV^2$ (see also Eq.\,(\ref{pcs})) of non-SC diborides
MeB$_2$ (Me=Zr, Nb, Ta) \cite{Naidyuk2} are shown. The cleanest
sample we have is ZrB$_2$ single crystal, and its PC EPI spectrum
demonstrates more pronounced features (see Fig.\,\ref{Heid}, left
panel). One recognizes a classical PC EPI spectrum from which one
can estimate the position of 3 main phonon peaks (for ZrB$_2$) and
obtain the lower limit of the EPI parameter $\lambda _{PC}\leq
0.1$ \cite{Naidyuk2}.

Essentially the similar spectra only with degradation of maxima
with bias rise were observed for another diborides, taking into
account their purity and increased EPI, which leads to the
transition from spectroscopic to non-spectroscopic (thermal)
regime of the current flow \cite{Naidyuk2}. The positions of the
low-energy peaks are proportional to the inverse square root of
the masses of $d$ metals \cite{Naidyuk2}, as expected. For NbB$_2$
and TaB$_2$ the phonon density of states (DoS) is measured by
means of neutron scattering \cite{Heid}. The position of phonon
peaks  corresponds to the PC spectra maxima (Fig.\,\ref{Heid}).
Because Nb and Zr have nearly the same atomic mass we suggest that
they should have similar phonon DoS.

\subsubsection{PC EPI spectra of MgB$_2$ in {\it c}-oriented films}

\begin{figure}
\includegraphics[width=10cm,angle=0]{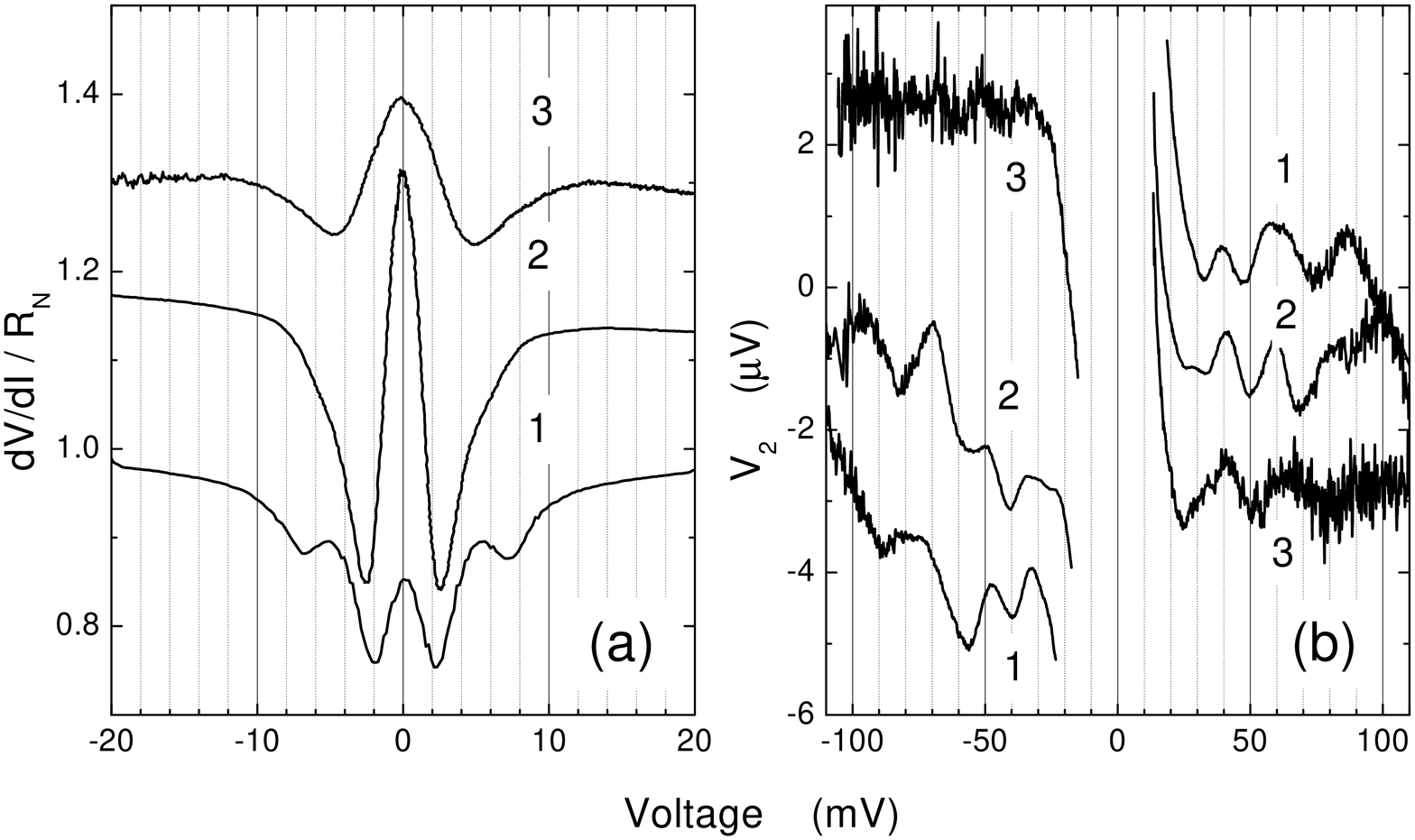}
 \vspace{-1cm}
\caption[]{Superconducting gap minima (left panel) and phonon
structure (right panel) in the spectra of thin film MgB$_2$-Ag PCs
with different resistances at $T=4.2 K$, $B=0$ with ($R_{0}$=45,
43, and 111 $\Omega $ for curves 1,2 and 3, respectively). The
modulation voltage $V_1$ at measuring of $V_2$ signal is 3.31,
2.78, and 2.5 mV, for curves 1,2 and 3, respectively. The numbers
of curves in (b) are the same as in (a). The curves in (a) are
offset relative to the bottom one for clarity.  After Yanson
et~al.~ \cite{YansonPRB}.} \label{Fig2phon}
\end{figure}

Unexpectedly, the stronger we suppress the superconductivity by
magnetic field or temperature in MgB$_2$, the less traces of
phonon structure remain in the $d^2V/dI^2$ derivative
\cite{YansonPRB}. This is in odd with the classical PCS, since the
{\it inelastic} phonon spectrum should not depend on the state of
electrodes in the first approximation (see section {\bf
Theoretical background of PCS}). Instead, most of the MgB$_2$
spectra in the SC state show reproducible structure in the phonon
energy range (Fig. \ref{Fig2phon}) which was not similar to the
expected phonon maxima superimposed on the rising background. This
structure disappears by transition to the normal state. Quite
interesting is that the intensity of this structure increases with
increase of the value of the small gap, which means that the gap
in $\pi $ band and observed phonon structure are connected
\cite{YansonPRB}. Based on the theoretical consideration mentioned
above, we conclude that the disorder in $\pi $ band is so strong
that it precludes to observe the {\it inelastic current}, and the
phonon non-linearities of excess current
\cite{YansonPRB,Om-Kul-Bel} play the main role, which does not
depend on the scattering.

Very rarely, we recovered (see \cite{Bobrov}) the structure in
$d^2V/dI^2$ which corresponds reasonably in shape to the phonon
DoS (above 30 meV). Thus, for this contact we assumed to observe
the {\it inelastic} PC spectrum for the $\pi$ band, which should
be compared to the Eliashberg EPI function for the same band
calculated in Ref. \cite{Golubov}. Both experimental spectrum and
$\pi$ band Eliashberg function do not show the anomalously high
intensity of $E_{2g}$ phonon mode, since only the Eliashberg
function for $\sigma$ band is the principal driving force for high
$T_c$ in MgB$_2$. The same conclusion should be ascribed to the
excess-current phonon structure, since it also corresponds to the
$\pi $ band. This band has much larger Fermi velocity and plasma
frequency along {\it c}-axis compared to $\sigma $ band
\cite{Brinkman}. Thus, in order to register the principal EPI with
$E_{2g}$ phonon mode the PC spectra  along {\it ab} plane should
be measured.

\subsubsection{PC EPI spectra in {\it ab}-direction}

\begin{figure}
\includegraphics[width=12cm,angle=0]{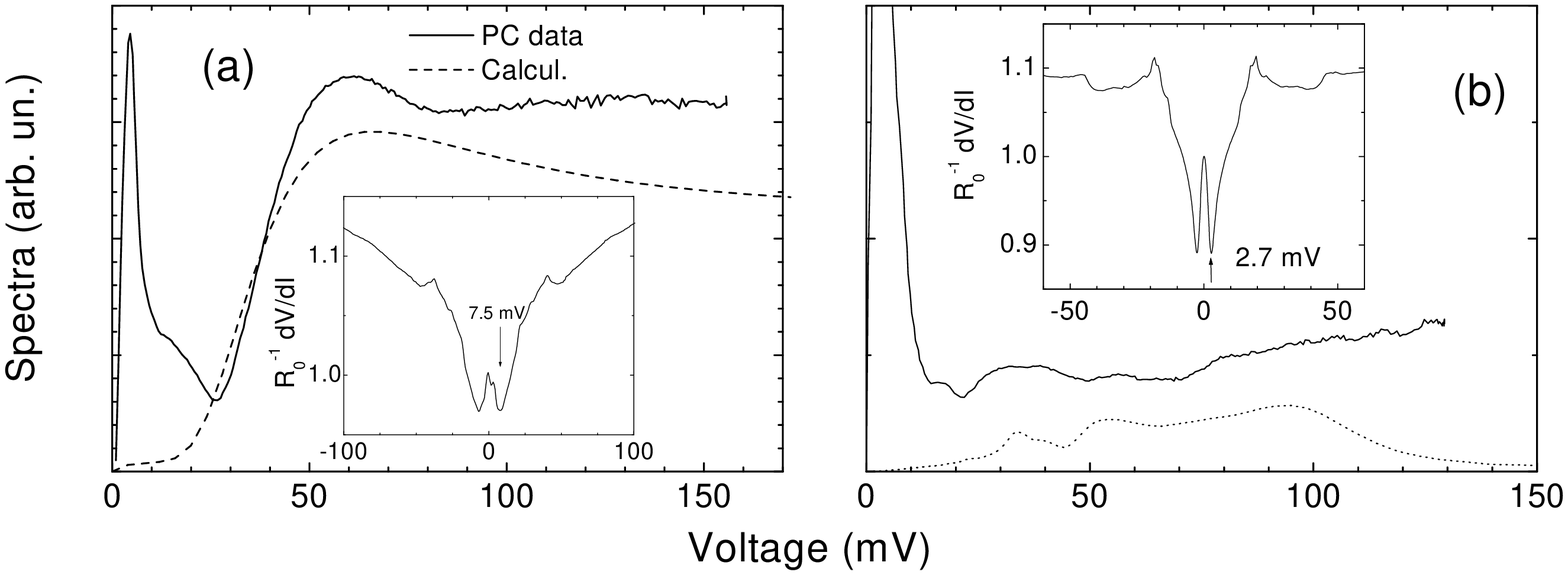}
 \vspace{-4cm}
\caption[]{(a) Comparison of the PC spectrum for $\sigma$ band
with the thermal spectrum for model spectral function as a
Lorentzian at 60 meV with a width of 2 meV (dashed line). Inset
shows large gap minima. (b) PC spectrum for $\pi$ band. Dashed
curve is the smeared theoretical Eliashberg function
\cite{Golubov}. Inset shows small gap minima. After Naidyuk
et~al.~ \cite{Naidyuk3}. } \label{MgB2scph}
\end{figure}

In Ref. \cite{Naidyuk3} PC EPI spectra for single crystal oriented
in {\it ab} plane were measured. As was mentioned above, the
nominal orientation of the contact axis to be parallel to {\it ab}
plane is not enough to be sure that this situation occurs in
reality. Moreover, even if one establishes the necessary
orientation (i.e., contact axis parallel to {\it ab} plane) the
spectra should reflect both bands with the prevalence of undesired
$\pi $ band, because due to spherical spreading of the current the
orientational selectivity of the metallic PC is much worse than
that for the plane tunnel junction, where it goes exponentially.
The large mixture of $\pi $-band contribution is clearly seen from
the gap structure in Fig. \ref{MgB2scph} (b), inset. Beyond the
wings at the biases corresponding to the large gap (supposed to
belong to $\sigma $-band gap) the deep minima located at the
smaller gap (correspondingly to $\pi $-band gap) are clearly seen.
The EPI spectrum of the same junction is shown in the main panel.
One can see that the non-linearities of the $I-V$ characteristic
at phonon biases are very small, and the reproducible structure
roughly corresponding to the Eliashberg EPI function of the $\pi $
band \cite{Dolgov,Golubov} appears in the bias range 20 $\div$ 60
mV. Above 60 mV PC spectrum broadens sufficiently sinking higher
lying phonon maxima. No remarkable contribution of $E_{2g}$ phonon
mode is observed, like a big maximum of EPI at $\approx 60\div
70$\,meV or a kink at $T\geq T_c$ for these biases.

Quite different spectrum is shown in Fig. \ref{MgB2scph}(a), which
is our key result. Consider first the $dV/dI(V)$ characteristics
(see inset). The energy gap structure shows gap minima
corresponding to the large gap ($ \sigma $-band gap). The increase
of $dV/dI(V)$ at larger biases is noticeably larger than in the
previous case.

Before the saturation at biases $\geq 100$ meV, where the phonon
DoS ends, the well resolved wide bump occurs in the PC spectrum,
which is located at about 60 meV.  Let us show that the bump is of
spectroscopic origin, that is the regime of the current flow
through the contact is not thermal, although the background at
large biases ($V\geq 100$ meV) is high. To do so, we compare this
bump with a PC spectrum in thermal regime for a model EPI
function, which consists of Lorentzian at 60\,meV with small
(2\,meV) width. Calculated according to Kulik \cite{Kulik_therm},
the thermal PC EPI spectrum, shown in Fig.\,\ref{MgB2scph}(a) as a
dashed line, is much broader.
Any further increase of the width of the model spectra will widen
the curve obtained. Comparing the experimental and model spectra
enables us to conclude, that in spite of big width the maximum of
experimental spectra is still correspond to the spectroscopic
regime.  Introducing greater disorder in the boron plane by
fabrication procedure or by trying another spots on the side-face
surface, the smeared thermal-like spectra were observed,
coinciding in shape with the dashed curve in
Fig.\,\ref{MgB2scph}(a), which together with corresponding
energy-gap structure can be ascribed to thermal limit mainly in
$\pi $ band, despite the bath temperature is low enough.

PC spectrum with broad maxima including also one at about 60 mV
were observed by \cite{Samuely} on polycrystalline MgB$_2$ samples
derived to the normal state by applying magnetic field and
increasing of the temperature.

The big width of the EPI peak connected with $E_{2g}$ phonon mode
is not surprising. Shukla {\it et al.} \cite{Shukla} measured the
phonon dispersion curves along $\Gamma $A and $ \Gamma $M
directions by means of inelastic X-ray scattering. The full width
at half maximum for $E_{2g}$ mode along $\Gamma $A direction
amounts to about 20-28 meV, which is well correspond to what we
observe in PC spectrum. If the phonon life time corresponds to
this (inverse) energy, then the phonon mean free path is about the
lattice constant \cite{Naidyuk3}, and due to phonon reabsorption
by nonequilibrium electrons, we should anticipate large background
in the PC spectra as observed.

For a contact with E$_{2g}$ phonon modes in
Fig.\,\ref{MgB2scph}(a) the nonlinearity of the $I-V$ curves due
to electron-phonon interaction can be estimated from the $dV/dI$
curves by about 10\%. This is comparable with nonlinearity
observed for non-SC diborides \cite{Naidyuk2} with small
electron-phonon coupling constant $\lambda _{PC}\leq 0.1$. The
reason of relatively low nonlinearity of $I-V$ curves and small
intensity of principal E$_{2g}$ phonon modes in spectra for
MgB$_2$ contacts can be the fact that anomalous strong interaction
is characteristic for the restricted group of phonons with
sufficiently small wave vector \cite{Mazin1}, whereas in PCS the
large angle scattering is underlined.

\section{Conclusions}

Comprehensive PCS investigations of $c$-axis oriented thin films
and single crystals of MgB$_2$ leads to the following conclusions:
\begin{itemize}

 \item The observed by Andreev reflection SC gaps in
MgB$_2$ are grouped at 2.4 and 7.0 meV and show basically a
BCS-like temperature dependence. The two gap structure merges
together in the case of strong elastic scattering remaining a
single gap at about 3.5 meV.

\item  Anomalous magnetic field dependencies of the gap structure
in PCs reflect peculiarity of the two band structure of the SC
order parameter in MgB$_2$. In particular, small gap survives up
to magnetic field close to the critical one for a large gap.

\item  The phonon structure in the PC spectra of MgB$_2$ can be
revealed by: i) the inelastic backscattering current, like for
ordinary PCS, and ii) by the energy dependence of the excess
current. They can be discriminated after destroying
superconductivity by magnetic filed or/and temperature, and
varying electron mean free path.

\item The prevailing appearance in the PC spectra of $E_{2g}$
boron mode, which mediates the creation of Cooper pairs, is seen
for PC with a large gap that is along $a-b$ direction in
accordance with the theory. The relatively small intensity of this
mode in the PC spectra is likely due to their small wave vector
and restricted phase volume.

\item  Related diborides (ZrB$_2$, NbB$_2$, and TaB$_2$) have PC
spectra proportional to the electron-phonon-interaction spectral
function, like in common metals and small EPI constant
corresponding to their non-SC state at helium temperature.
\end{itemize}

\section*{Acknowledgements}
The authors are grateful to N. L. Bobrov, P. N. Chubov, V. V.
Fisun, O. E. Kvitnitskaya, L. V. Tyutrina for collaboration during
MgB$_2$ investigation and S.-I. Lee and S. Lee for samples
providing. The work in Ukraine was supported by the State
Foundation of Fundamental Research under Grant $\Phi$7/528-2001.

\end{document}